# Tuning carrier type and density in $Bi_2Se_3$ by Ca-doping

Zhiyong Wang,[1] Tao Lin,[1] Peng Wei,[1] Xinfei Liu,[1] Randy Dumas,[2] Kai Liu,[2] and Jing Shi[1]

[1]Department of Physics and Astronomy, University of California, Riverside, CA 92506

[2]Department of Physics, University of California, Davis, CA 95616

Abstract

The carrier type and density in $Bi_2Se_3$ single crystals are systematically tuned by introducing a calcium (Ca) dopant. A carrier density of ~$1\times10^{17}$ $cm^{-3}$ which corresponds to ~25 meV in the Fermi energy is obtained in both *n*- and *p*-type materials. Electrical transport properties show that the insulating behavior is achieved in low carrier density crystals. In addition, both the band gap and reduced effective mass of carriers are determined.

Among several topological insulator candidates, $Bi_2Se_3$ is predicted to have the largest band gap of ~ 0.3 eV.[1] Recent angle-resolved photoemission spectroscopy (ARPES) studies confirmed many predicted characteristics of its bulk and surface electronic band structures.[2,3] At the exact stoichiometry, however, $Bi_2Se_3$ does not exhibit the expected insulating behavior due to excess charge carriers caused by selenium vacancies. By introducing dopants such as calcium (Ca), it was shown by ARPES that the Fermi level could be moved to the bulk band gap. An insulating behavior was observed in electrical resistivity measurements of Ca-doped materials. In order to explore the exotic nature of the surface metallic states, it is very important to *control* the position of the Fermi level, and consequently the carrier density in the bulk bands.

In this work, we systematically vary the Ca-concentration in order to tune the carrier density and the carrier type in high-quality $Bi_2Se_3$ single crystals. We are able to dope the material very close to the compensation point so that the insulating transport behavior is obtained. From the optical and electrical transport measurements, we determine the band gap of $Bi_2Se_3$ and the reduced effective mass of the electrons and holes.

Single crystals of $Ca_xBi_{2-x}Se_3$ ($x$=0, 0.005, 0.012, 0.015, and 0.020) were grown using a multi-step heating method described previously.[2, 4, 5, 6] First, high-purity $Bi_2Se_3$ (99.999%) compound and Ca (99.98%) were mixed according to the stoichiometry and sealed in an evacuated quartz tube. The tube was then heated to and kept at 800ºC for 24 hours in a programmable furnace. It was subsequently cooled to and kept at 500ºC for 72 hours before it was finally cooled down to room temperature.

Single crystals up to ~1 inch in length and 1/4 inch in diameter can be grown by this method. Structural characterizations have been performed with a Bruker D8 4-circle thin film x-ray diffractometer on large crystals. The $\theta/2\theta$ x-ray diffraction pattern of an undoped ($x$=0) $Bi_2Se_3$ crystal is shown in Fig. 1. Only the (003) family of diffraction peaks are observed, indicating that the crystals are exclusively trigonal-axis oriented. The full width at half maximum (FWHM) of the (006) peak is less than 0.037°. That of the rocking curve measured for the (006) diffraction peak is less than 0.039°, indicating high crystal quality and large in-plane coherence length. The inset shows an azimuthal $\Phi$ scan measured at a title angle of $\Psi$=56.4°, exhibiting the expected threefold symmetry of the (015) diffraction plane. Similar $\Phi$ scans show that the (116), (101), and (110) peaks exhibit the expected sixfold, threefold, and sixfold symmetry, respectively. These results confirm that the $Bi_2Se_3$ samples are indeed high quality single crystals.[7]

The crystals can be easily cleaved as shown in the inset of Fig. 2. Thin flakes can be prepared by repeated cleaving for both transport and Fourier-transform infrared (FTIR) transmittance and reflectance measurements. For transmittance measurements, the thin flakes are typically ~ 10-100 μm thick. The onset of the infrared (IR) transmittance on the high-frequency side of the FTIR spectra, which marks the inter-band transition, is determined for different samples. For direct comparison, the exact same samples are then used for the Hall measurement to determine the carrier type and carrier concentration. Both resistivity and the Hall coefficient measurements were performed in a closed-cycle system using the Van der Pauw method.

The transmittance, $T$, of various samples is shown in Fig. 2. At low wave-numbers (~ 500 $cm^{-1}$), $T$ rapidly decreases as the cut-off in detector's sensitivity is approached. At high wave-numbers (> 2,700 $cm^{-1}$), $T$ vanishes as the absorption is turned on due to the inter-band transition. For the intermediate spectral range, $T$ is finite, indicative of the band gap of semiconductor materials. The oscillations in $T$ are a result of the Fabry-Perot interference of IR between the two flat and specular surfaces of the samples. The interference fringes serve as a convenient way of measuring the thickness of the flakes for three-dimensional carrier density ($n_{3D}$) determination. We define the onset of the inter-band transition as $E_o$, which depends on the position of the Fermi level, the band gap $E_g$, and the effective mass of the carriers. For the undoped sample or $x$=0, $E_0$ is ~2,400 $cm^{-1}$ or 0.30 eV, which is the highest among all five samples. For $n$-type materials, $E_0$ corresponds to the inter-band electronic transition from the valence band to the Fermi level in the conduction band. The initial and final states are connected by the same wave-vector according to the selection rule. The $E_0$ decreases as the electron density is reduced. This decrease occurs as more Ca atoms are incorporated into $Bi_2Se_3$. The decreasing trend continues until $x$ reaches 0.012, where $E_0$ then begins to rise as the compensation point is passed and the Fermi level now moves to the valence band. As the hole concentration increases further, $E_0$ increases because the position of the Fermi level sinks downward in the valence band.

To determine the carrier density, $n_{3D}$, for different Ca-doping levels, we have measured the Hall coefficient on the same thin samples used for transmittance measurements. The undoped $Bi_2Se_3$ sample is $n$-type and $n_{3D}$ is ~ $5\times10^{18}$ $cm^{-3}$. For $x$=0.012, which has the lowest inter-band transition energy, both FTIR and Hall coefficient measurements are carried out on multiple flakes cleaved from different parts of the same bulk crystal. Although $E_0$ stays approximately the same, the carrier type changes from sample to sample. Quantitative Hall coefficient measurements show that the lowest carrier density is ~ $4\times10^{17}$ $cm^{-3}$, and the sample-to-sample density variation is ~ $5\times10^{17}$ $cm^{-3}$. The variation in both carrier type and $n_{3D}$ indicates some degree of spatial variations in Ca concentration in bulk crystals, which can be more easily detected near the compensation point.

Fig. 3 shows the relationship between $E_0$ and $n_{3D}$ for samples with different Ca-doping levels. For both $n$- and $p$-types, the inter-band transition is related to the Fermi wave-vector $k_F$ or consequently $n_{3D}$, the band gap $E_g$, and the reduced effective mass $\mu^* = \frac{m_C^* m_V^*}{m_C^* + m_V^*}$, i.e. $E_0 = \hbar\omega = E_g + E_C + E_V = E_g + \frac{\hbar^2 k_F^2}{2\mu^*} = E_g + \frac{\hbar^2 (3\pi^2 n_{3D})^{2/3}}{2\mu^*}$

From Fig. 3, we can readily extract both $E_g$ and $\mu^*$. The best fit of $E_0 = A n_{3D}^{2/3} + \mathrm{B}$ to our data gives $E_g$ = 0.093 eV and $\mu^*$ = 0.072 $m_e$. ($m_e$: free electron mass). The band gap value is smaller than the theoretical value of 0.3 eV computed for T = 0 K[1] or the low temperature value observed by ARPES,[8] but is reasonably close to the values determined by other bulk sensitive techniques[9,10] (e.g. 115 meV in ref. 9). Although it is not possible to determine the effective masses for both electrons and holes independently, the reduced effective mass obtained from our experiments agrees with the value calculated from the electron and hole masses reported in published works. [10,11] Using the estimated effective electron mass ($m_e^*$=0.14 $m_e$) and the dielectric constant ($\varepsilon$=3), we calculated the plasma frequency for a high-density *n*-type sample ($x$=0.5% and $n_{3D}$=7.2×10$^{18}$ cm$^{-3}$) to be ~ 700 cm$^{-1}$, which agrees well with the position of the plasma dip in the measured reflectance spectrum (~ 611 cm$^{-1}$).

The carrier density spans over an order of magnitude in this set of samples. For the lowest density samples, the Fermi energy is ~ 25 meV above (below) the top (bottom) of the valence (conduction) band. Therefore, it is expected that those low-density samples have pronounced temperature dependence in their electrical transport properties. We have performed electrical resistivity measurements on 5 samples. The electrical resistivity varies significantly for samples with different carrier densities, spanning over two orders of magnitude at room temperature. To show the temperature dependence of all samples, the data are displayed with a semi-log plot in Fig. 4 and the normalized resistivity is shown in the inset. The two most resistive samples are both originally cleaved from the $x$=0.012 crystal, and the carrier density $n_{3D}$ is ~ +7.1×10$^{17}$ cm$^{-3}$ (*p*-type) and ~ -1.6×10$^{18}$ cm$^{-3}$ (*n*-type), respectively. Clearly, the three samples with low room temperature resistivity ($x$=0, 0.005, and 0.020) show metallic behavior with a positive temperature coefficient of resistivity, i.e. $d\rho/dT > 0$; on the other hand, the *n*-type $x$=0.012 sample shows an insulating behavior, i.e. $d\rho/dT < 0$; while the *p*-type $x$=0.012 sample shows a relatively flat temperature dependence of resistivity. We attribute such drastically different temperature dependences to the carrier density variation as the temperature is lowered.

In conclusion, we have successfully tuned both the carrier type and density by incorporating calcium into $Bi_2Se_3$ single crystals. From the inter-band transition energy measured by FTIR and carrier density measured by the Hall coefficient, we have determined the band gap of $Bi_2Se_3$ and the reduced effective mass. By tuning the carrier density close to the compensation point, we have demonstrated the insulating behavior in low density bulk crystals, which paves the way for studying the topological insulator properties in such materials.

We thank A. LaForge and D. Basov for their help and discussions. Work at UCR was supported in part by DOE DE-FG02-07ER46351 and NSF (ECCS-0802214). Work at UCD was supported in part by NSF (ECCS-0725902, ECCS-0925626). K.L. acknowledges support from a UCD Chancellor's Fellowship.

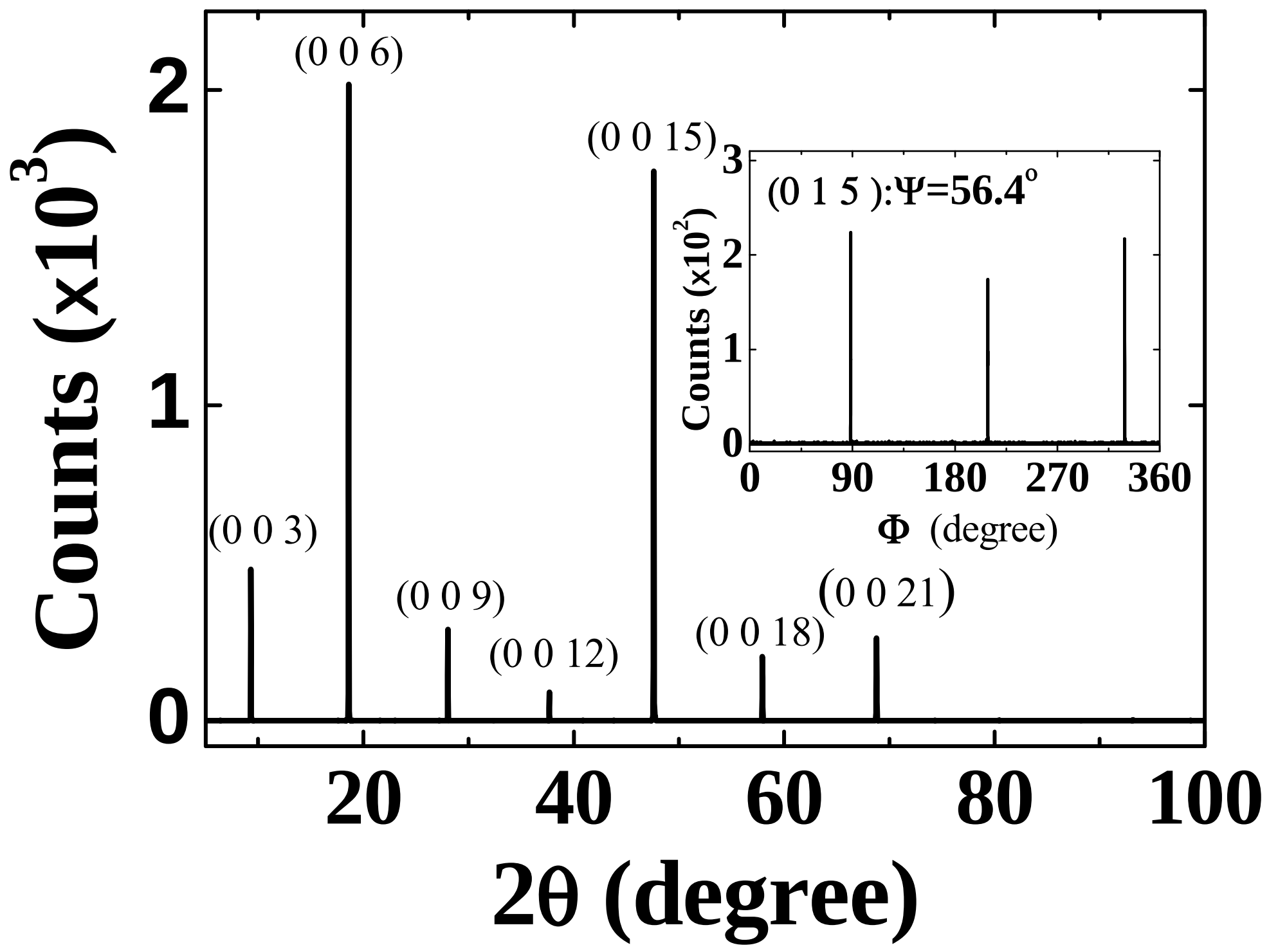

Fig. 1. X-ray diffraction pattern of an undoped ($x$=0) $Bi_2Se_3$ crystal. The inset is an azimuthal $\Phi$ scan, exhibiting the expected threefold symmetry of the (015) diffraction peak.

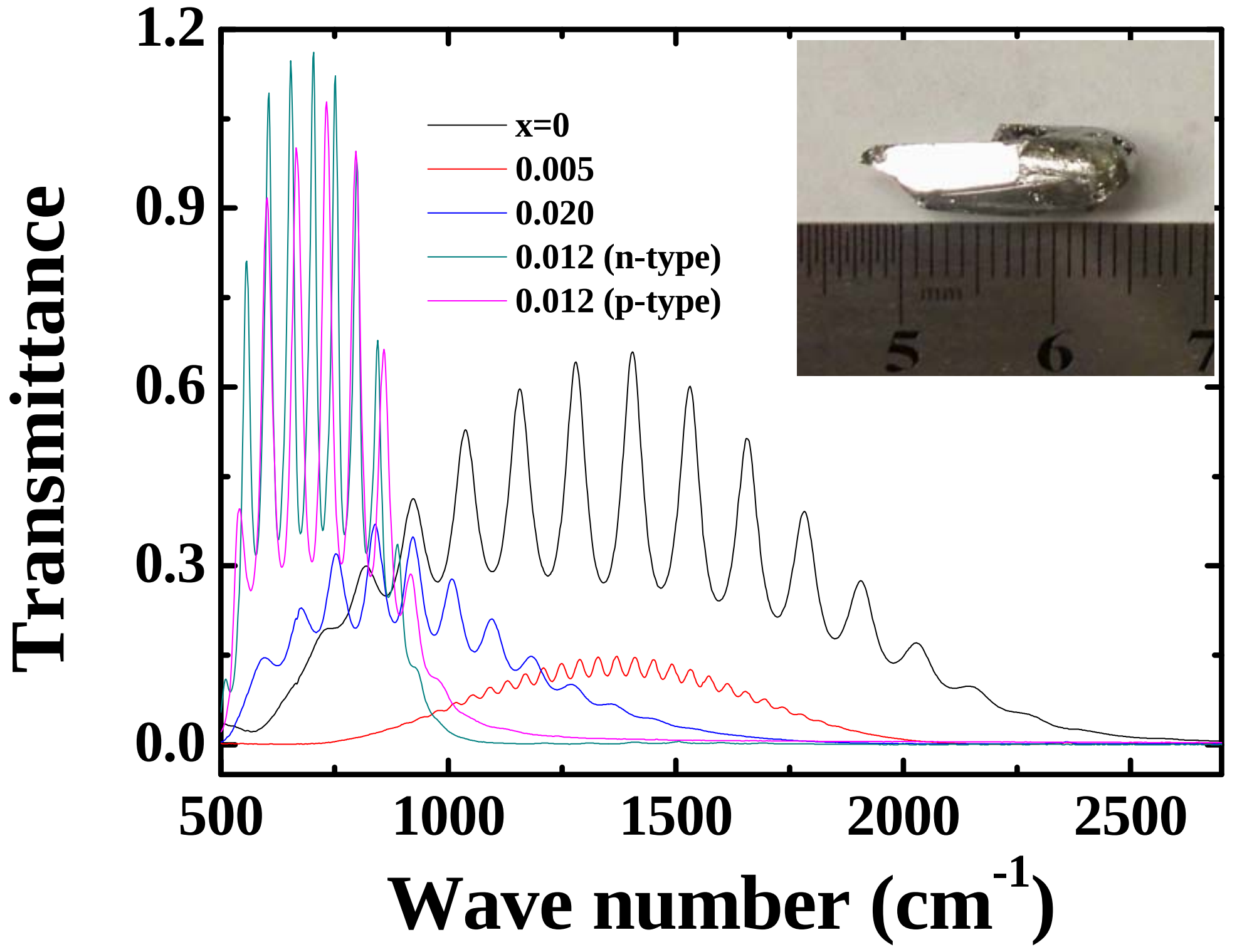

Fig. 2. FTIR transmittance spectra for both doped and undoped samples. Two $x$=0.012 Ca-doped samples have the smallest inter-band transition energy, $E_0$. Samples with higher (p-type) and lower (n-type) Ca doping have a larger $E_0$. This indicates that $x$=0.012 Ca is close to the compensate point. The inset is a picture of a cleaved Ca-doped $Bi_2Se_3$ crystal.

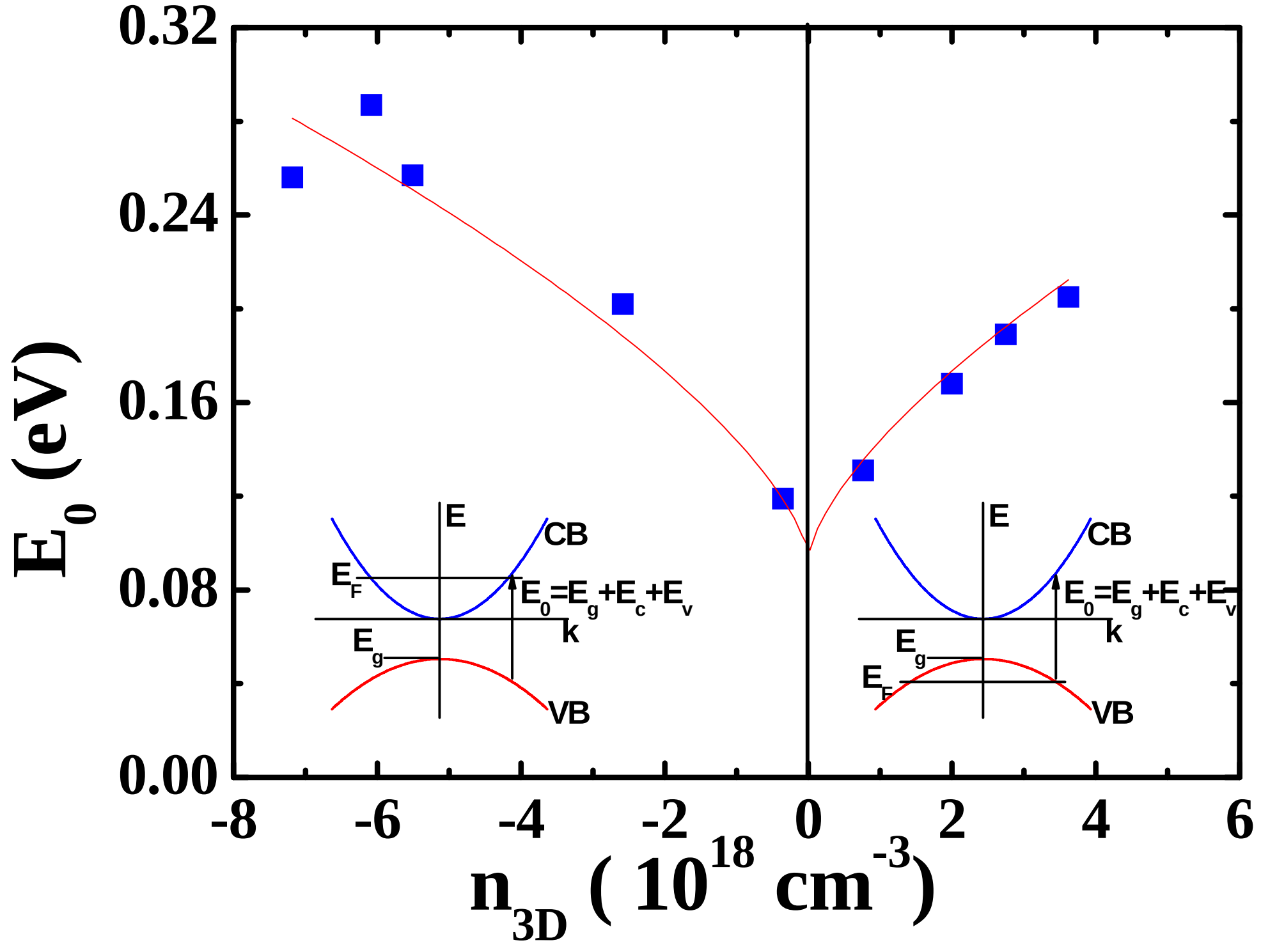

Fig. 3. Inter-band transition energy, $E_0$ vs. 3D carrier concentration, $n_{3D}$. Negative values in $n_{3D}$ represent *n*-type while positive values represent *p*-type. The best fit of $E_0 = An_{3D}^{2/3} + B$ to the data results in $E_g = B = 0.093$ eV and the reduced effective mass, $\mu^* = 0.072\ m_e$. The insets show the sketches of inter-band transitions for *n*- and *p*-type samples. We use the same fitting parameters for both sides.

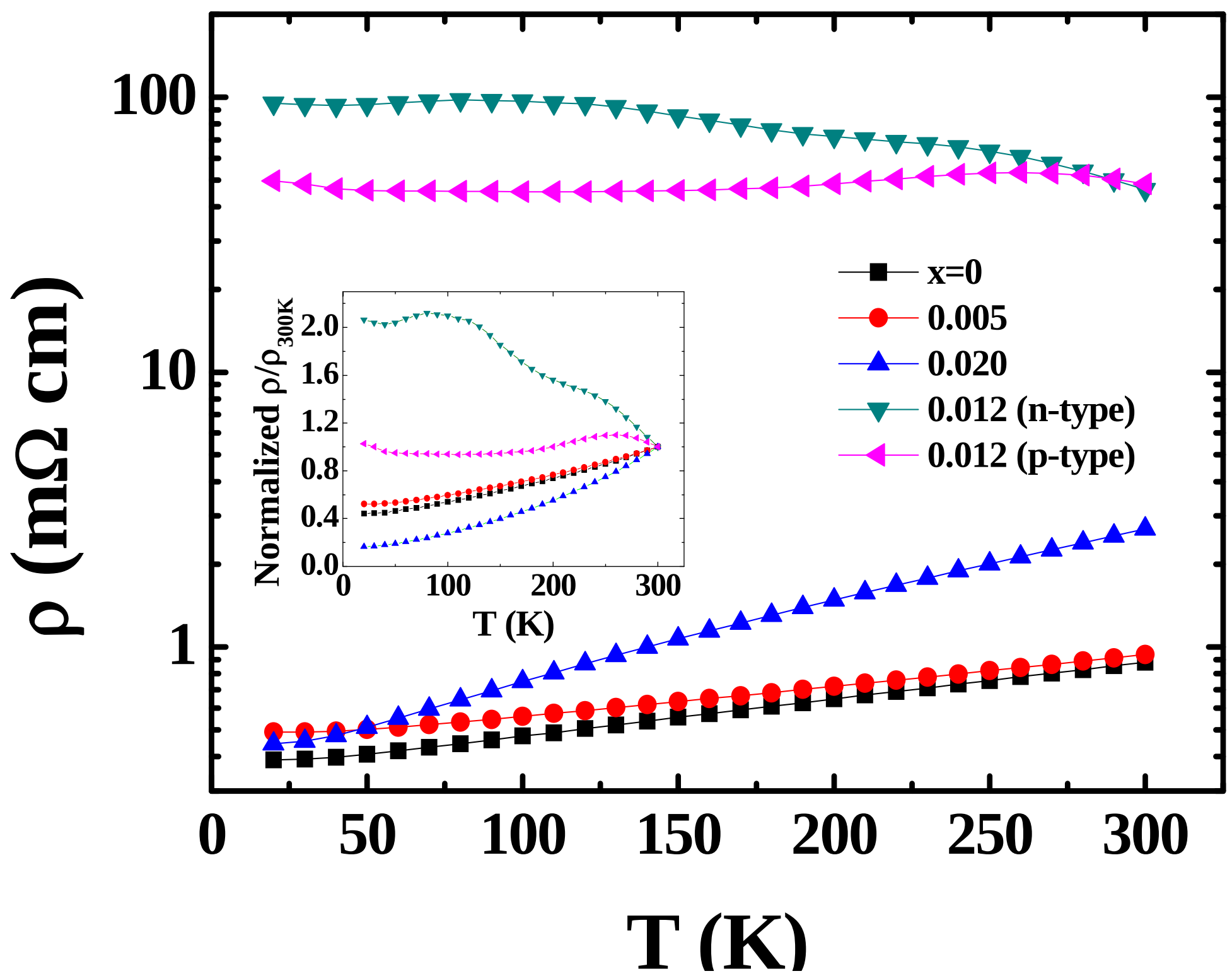

Fig. 4 Temperature-dependent resistivity $\rho$ curves for both doped and undoped samples. One *n*-type $x$=0.012 sample shows the insulating behavior. Its resistivity is two orders of magnitude greater than that of the metallic samples. The inset is the normalized resistivity $\rho/\rho_{300K}$ vs. $T$.